\documentclass[journal]{IEEEtran}

\usepackage{url}
  \usepackage[pdftex]{graphicx}
\usepackage{comment}
\usepackage[font=small,hang]{caption}
\usepackage{amsmath,amssymb} 
\usepackage{color,soul}
\usepackage{float}
\usepackage{booktabs}
\usepackage{multirow}
	\usepackage{textcomp}
\usepackage{array}
\usepackage[comma,sort&compress,numbers]{natbib}
\usepackage[font=small]{caption}
\newcolumntype{P}[1]{>{\centering\arraybackslash}p{#1}}
\newcolumntype{M}[1]{>{\centering\arraybackslash}m{#1}}
\usepackage{url}
\usepackage{siunitx}
\usepackage{scrextend}
   \graphicspath{{./figures/}}
\usepackage{hyperref}
\begin{document}

\title{BVI-DVC: A Training Database for Deep Video Compression}

\author{Di Ma, Fan Zhang,~\IEEEmembership{Member,~IEEE,}
        and David R. Bull,~\IEEEmembership{Fellow,~IEEE}
\thanks{D. Ma, F. Zhang and D. R. Bull are with the Department of Electrical and Electronic Engineering, University of Bristol, Bristol, BS8 1UB, UK (e-mail: di.ma@bristol.ac.uk; fan.zhang@bristol.ac.uk; dave.bull@bristol.ac.uk).}

\thanks{}}


\maketitle

\begin{abstract}
Deep learning methods are increasingly being applied in the optimisation of video compression algorithms and can achieve significantly enhanced coding gains, compared to conventional approaches. Such approaches often employ Convolutional Neural Networks (CNNs) which are trained on databases with relatively limited content coverage. In this paper, a new extensive and representative video database, BVI-DVC\footnote{\label{fn:db}The BVI-DVC database can be downloaded from: \url{https://vilab.blogs.bristol.ac.uk/?p=2375}.}, is presented for training CNN-based video compression systems,  with specific emphasis on machine learning tools that enhance conventional coding architectures, including spatial resolution and bit depth up-sampling, post-processing and in-loop filtering. BVI-DVC contains 800 sequences at various spatial resolutions from 270p to 2160p and has been evaluated on ten existing network architectures for four different coding tools. Experimental results show that this database produces significant improvements in terms of coding gains over three existing (commonly used) image/video training databases under the same training and evaluation configurations. The overall additional coding improvements by using the proposed database for all tested coding modules and CNN architectures are up to 10.3\% based on the assessment of PSNR and 8.1\% based on VMAF.
\end{abstract}


\IEEEpeerreviewmaketitle

\section{Introduction}

\IEEEPARstart{F}{rom} the introduction of the first international standard in 1984, video compression has played an essential role in the application and uptake of video technologies across film, television, terrestrial and satellite transmission, surveillance and particularly Internet video \cite{bull2014communicating}. Inspired by recent breakthroughs in AI technology, deep learning methods such as Convolutional Neural Networks (CNNs) have been increasingly exploited into video coding algorithms \cite{ma2019image} to provide significant coding gains compared to conventional approaches based on classic signal/image processing theory. 

It is noted that these learning-based compression approaches demand volumes of training material much greater than typically used for conventional compression or existing machine learning methods. These should include diverse content covering different formats and video texture types. As far as we are aware, no such public data sets for this purpose currently exist, and most learning-based coding methods
\cite{liu2020deep} have been currently trained on image or video databases, which were mainly designed for computer vision applications, e.g. super-resolution. Most of these databases do not provide sufficient content coverage and diversity. As a result, the generalisation of networks cannot be ensured in the context of video coding, and the optimum performance of employed CNN networks has not been achieved when trained on these databases. 

In this paper, a new video database, referred as BVI-DVC, is proposed for training CNN-based video coding algorithms, in particular those tools that enhance the performance of conventional compression algorithms. These include spatial resolution re-sampling, bit depth  re-sampling, post-processing and in-loop filtering, all of which have achieved significant coding gains compared to other deep learning enhancements. BVI-DVC contains 800  progressive-scanned video clips at a wide range of spatial resolutions from 270p to 2160p, with diverse and representative content. To demonstrate its training effectiveness, compared to three commonly used training databases \cite{Agustsson_2017_CVPR_Workshops,nah2019ntireData,liu2017robust}, BVI-DVC has been utilised to train ten popular CNN architectures \cite{dong2015image,dong2016accelerating,kim2016accurate,tai2017image,lim2017enhanced,ledig2017photo,wang2018esrgan,zhang2018image,zhang2018residual,ma2019perceptually} using four video coding tools: post-processing (PP),  in-loop filtering (ILF), spatial resolution adaptation (SRA) and effective bit depth adaptation (EBDA). The resulting CNN-based coding tools were then integrated into the Test Model (HM 16.20) of the current High Efficiency Video Coding (HEVC) standard \cite{hevc}, and evaluated based on the Joint Video Experts Team (JVET) Common Test Conditions (CTC) \cite{jvetctc}. 

The primary contributions of this paper are:

\begin{enumerate}
    \item A publicly available video database (BVI-DVC) specifically developed for training deep video coding compression algorithms.
    \item An analysis of the performance of BVI-DVC in the context of four compression tools compared with three existing (commonly used) databases.
    \item A performance comparison of ten popular CNN architectures based on identical training materials and evaluation configurations.
\end{enumerate}

The remainder of the paper is structured as follows. Section II summarises the related work in deep learning-based video compression and commonly used image/video training databases. Section III introduces the proposed (BVI-DVC) video training database, while Section IV presents the experimental configurations employed for evaluating the effectiveness of the proposed database. The results and discussions are provided in Section V, and Section VI concludes the paper and presents some suggestions for future work.  

\section{Background}
\label{sec:review}

With increasing spatial resolutions, higher frame rates, greater dynamic range and the requirement for multiple viewpoints, accompanied by dramatic increases in user numbers, video
content has become the primary driver for increased internet bandwidth. How we represent and communicate video (via compression) is key in ensuring that the content is delivered at an appropriate quality, while maintaining compatibility with the transmission bandwidth.

To address this issue, ISO (MPEG) and ITU initiated the development of a new coding standard, Versatile Video Coding (VVC) \cite{bross2018versatile} in 2018, targeting increased coding gain (by 30-50\%) compared to the current HEVC standard. Concurrently, the Alliance for Open Media (AOM - formed in 2015) finalised an open-source, royalty-free media delivery solution (AV1) \cite{av1}  in 2018, offering performance competitive with HEVC \cite{katsenou2019subjective}.

\subsection{Machine learning based compression} Image and video compression based on deep neural networks has become a popular research topic offering evident enhancements over conventional coding tools for: intra prediction \cite{yeh2018hevc,li2018fully}, motion estimation \cite{zhao2018enhanced,zhao2019enhanced}, transforms \cite{puri2017cnn,jimbo2018deep}, quantisation \cite{alam2015perceptual}, entropy coding \cite{song2017neural,ma2018convolutional}, post-processing \cite{li2017cnn,zhao2019cnn} and loop filtering \cite{jia2019content,ma2020mfrnet}. New coding tools such as format adaptation \cite{lin2018convolutional,zhang2019vistra2,ma2020gan} and virtual reference frame optimisation \cite{zhao2019enhanced} have also been reported. Other work has implemented a complete coding framework based on neural networks using end-to-end training and optimisation \cite{balle2016end,rippel2019learned,lu2019dvc,djelouah2019neural,habibian2019video}. It is noted that, despite their performance benefits, few of these approaches are currently being adopted by the latest VVC video coding standard. This is due to the high computational complexity and the large GPU memory requirements associated with CNN computation.

\subsection{Training Databases} 
Training databases are a critical component for optimising the performance of machine learning based algorithms. A well designed training database can ensure good model generalisation and avoid potential over-fitting problems \cite{Zhu2016,tian2018tdan}. As far as we are aware, there is no publicly available database which is specifically designed for learning-based video coding. Researchers, to date, have typically employed training databases developed for other purposes (such as super-resolution, frame interpolation and classification) for training. Notable publicly available image and video training databases are summarised below. 

\begin{itemize}
    \item \textbf{ImageNet} \cite{russakovsky2015imagenet} is a large image database primarily designed for visual object recognition. It contains more than 14 million RGB images at various spatial resolutions (up to 2848p) covering a wide range of natural content. It has also been used as a training database for single image super-resolution \cite{ledig2017photo}. 
    
    \item \textbf{DIV2K} \cite{Agustsson_2017_CVPR_Workshops} contains 1000 RGB source images with a variety of content types, which was firstly developed for super-resolution. It has currently been employed as training material by several JVET proposals \cite{r:JVETO0079,JVET0158} and many other CNN-based coding algorithms \cite{wang2018denseInLoop,wang2019partition}. 
    
    \item \textbf{BSDS} \cite{martin2001database} is an image database originally developed for image segmentation. It contains 500 RGB images, and has been used to train CNN-based loop filters \cite{jia2019content} for video coding. Comparing to DIV2K, BSDS has fewer source images and lower spatial resolution (481$\times$321).
    
    \item \textbf{Vimeo} \cite{xue2019video} is a video database originally developed for training CNN-based optical flow and temporal super-resolution approaches. It contains 89,800 sequences at spatial resolutions up to 448$\times$256. A constraint is imposed on motion vector magnitudes between any two adjacent frames and content with dynamic textures has not been included in this database. Vimeo has not been frequently employed for deep learning based coding approaches and in particular has not been used for those approaches that exhibit superior improvements over standard video codecs (e.g. HEVC and VVC) \cite{liu2020deep}.
    
    \item \textbf{CD} (Combined Database) in \cite{liu2017robust} is a video database combining source content from the LIVE Video Quality Assessment Database \cite{seshadrinathan2010study}, MCL-V Database \cite{lin2015mcl} and TUM 1080p Database \cite{keimel2010visual} and has been employed to train CNN-based super-resolution approaches \cite{liu2017robust}. It contains 29 sequences at two different spatial resolutions, 1920$\times$1080 and 768$\times$432. 
    
    \item \textbf{REDS} \cite{nah2019ntireData} is a video database developed for training video super-resolution algorithms \cite{nah2019ntire}, which contains 300 video clips with spatial resolution 1280$\times$720.
    
    \item \textbf{UCF101} \cite{soomro2012ucf101} is a large video training database initially designed for human action recognition, and has been frequently used for training CNN-based temporal frame interpolation and motion prediction approaches \cite{liu2017videodvf,szeto2019temporally,zhang2018frame}. It contains 13320 videos collected from YouTube, which consist of 101 types of human actions. All the sequences in UCF-101 have a relatively low spatial resolution of 320$\times$240. 
\end{itemize}

Modern video coding algorithms are required to process content with diverse texture types at high spatial resolutions and bit depths. For example, the standard test sequences included in the JVET Common Test Conditions (CTC) dataset include video clips at UHD resolution (2160p) at a bit depth of 10 bits, with various static and dynamic textures\footnote{\label{fn:texture} Here we follow the definition of textures in \cite{zhang2011parametric,ma2019synthetic}. Static textures are associated with rigid patterns undergoing simple movement or subject to camera movement, while dynamic textures have complex and irregular movements, e.g. water, fire or steam.}. However none of the training databases mentioned above contain image or video content with high spatial resolution and bit depth, and most do not include any dynamic texture content. 

\section{The BVI-DVC Video Training Database}
\label{sec:bvidl}

\begin{table*}[htbp]
\centering
\caption{Key features of eight training databases including BVI-DVC.}
\begin{tabular}{l | r | r| r |r |r| r|r|r}
\toprule
Features & ImageNet \cite{russakovsky2015imagenet} & DIV2K \cite{Agustsson_2017_CVPR_Workshops} & BSDS
\cite{martin2001database} & Vimeo
\cite{xue2019video} & CD \cite{liu2017robust} & REDS \cite{nah2019ntireData} &UCF101 \cite{soomro2012ucf101} & BVI-DVC\\
\midrule
Image or Video? &Image &Image &Image &Video & Video & Video &Video & Video \\
\midrule
Seq Number &14M & 1000& 500 &89,800& 29 & 300 & 13,320  & 800 \\
\midrule
Max Resolution &2848p & 1152p & 321p &256p&1080p & 720p & 240p & 2160p \\
\midrule
Bit depth &8 &8 &8 &8 &8 & 8 &8 & 10\\
\midrule
Various textures? &No&No&No&No &No & No &No & Yes \\
\bottomrule
\end{tabular}
\label{tab:feature}
\end{table*}

\begin{figure*}[ht]
\centering
\scriptsize
\centering
\begin{minipage}[b]{0.19\linewidth}
\centering
\centerline{\includegraphics[width=1.01\linewidth]{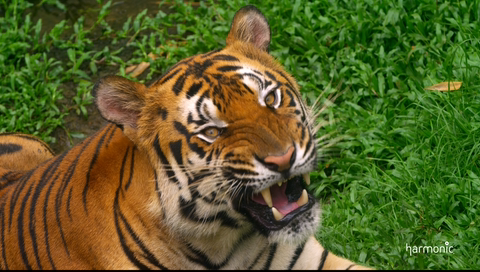}}
(a) Animal
\end{minipage}
\begin{minipage}[b]{0.19\linewidth}
\centering
\centerline{\includegraphics[width=1.01\linewidth]{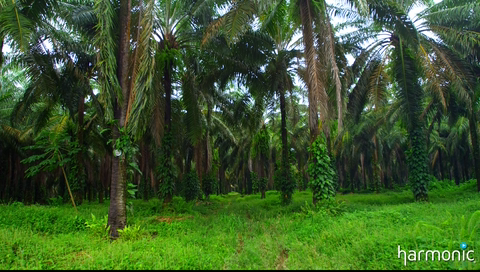}}
(b) Wood
\end{minipage}
\begin{minipage}[b]{0.19\linewidth}
\centering
\centerline{\includegraphics[width=1.01\linewidth]{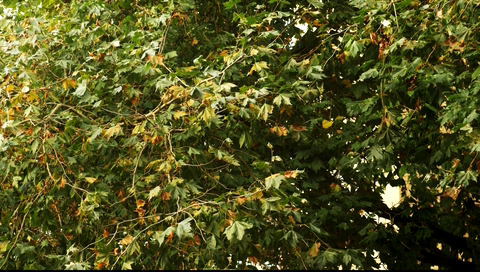}}
(c) Leaves
\end{minipage}
\begin{minipage}[b]{0.19\linewidth}
\centering
\centerline{\includegraphics[width=1.01\linewidth]{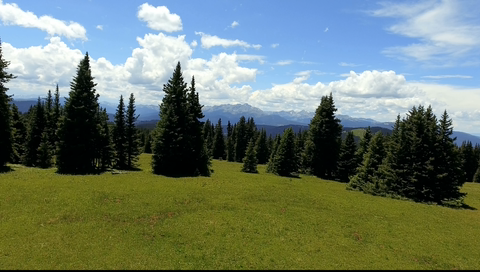}}
(d) Mountain
\end{minipage}
\begin{minipage}[b]{0.19\linewidth}
\centering
\centerline{\includegraphics[width=1.01\linewidth]{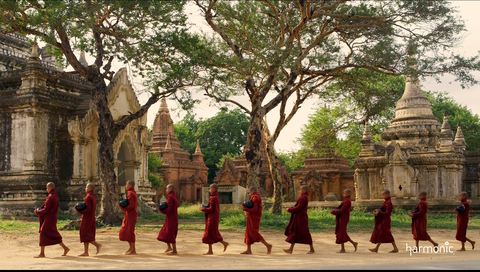}}
(e) Myanmar
\end{minipage}

\begin{minipage}[b]{0.19\linewidth}
\centering
\centerline{\includegraphics[width=1.01\linewidth]{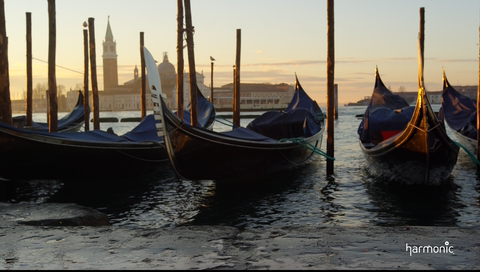}}
(f) Venice
\end{minipage}
\begin{minipage}[b]{0.19\linewidth}
\centering
\centerline{\includegraphics[width=1.01\linewidth]{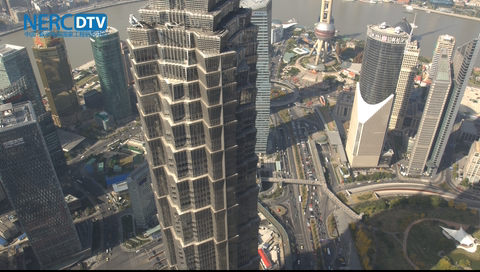}}
(g) Tall Buildings
\end{minipage}
\begin{minipage}[b]{0.19\linewidth}
\centering
\centerline{\includegraphics[width=1.01\linewidth]{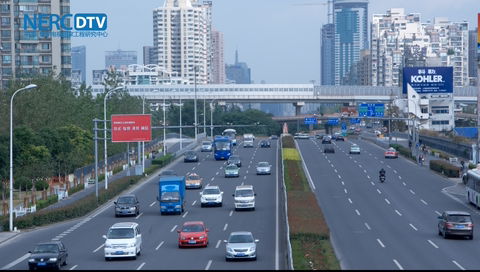}}
(h) Traffics
\end{minipage}
\begin{minipage}[b]{0.19\linewidth}
\centering
\centerline{\includegraphics[width=1.01\linewidth]{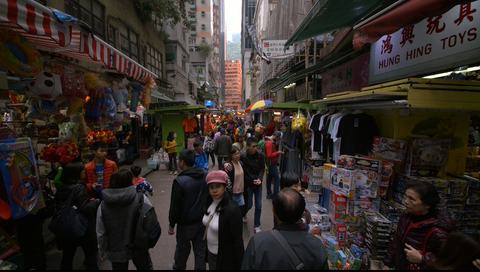}}
(i) Market
\end{minipage}
\begin{minipage}[b]{0.19\linewidth}
\centering
\centerline{\includegraphics[width=1.01\linewidth]{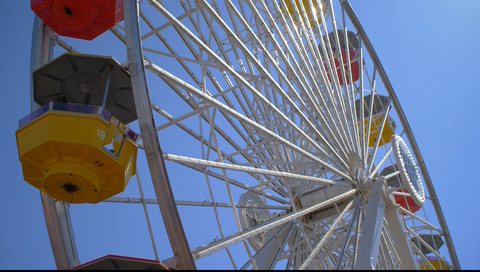}}
(j) Ferris Wheel
\end{minipage}

\begin{minipage}[b]{0.19\linewidth}
\centering
\centerline{\includegraphics[width=1.01\linewidth]{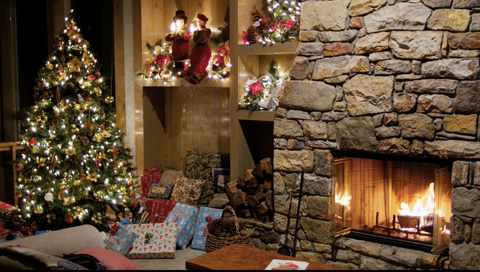}}
(k) Room
\end{minipage}
\begin{minipage}[b]{0.19\linewidth}
\centering
\centerline{\includegraphics[width=1.01\linewidth]{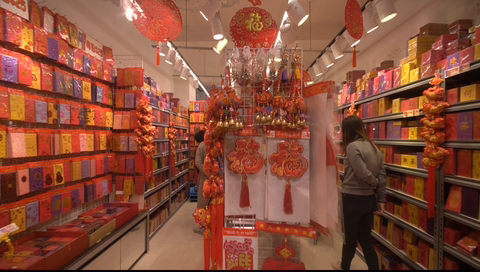}}
(l) Store
\end{minipage}
\begin{minipage}[b]{0.19\linewidth}
\centering
\centerline{\includegraphics[width=1.01\linewidth]{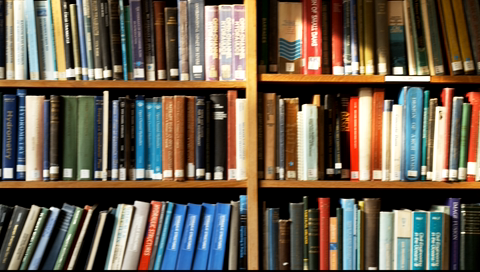}}
(m) Bookcase
\end{minipage}
\begin{minipage}[b]{0.19\linewidth}
\centering
\centerline{\includegraphics[width=1.01\linewidth]{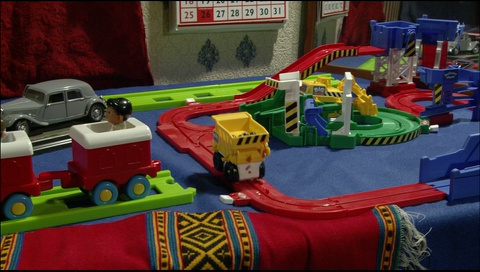}}
(n) Toy
\end{minipage}
\begin{minipage}[b]{0.19\linewidth}
\centering
\centerline{\includegraphics[width=1.01\linewidth]{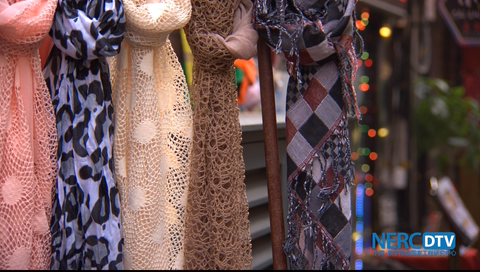}}
(o) Scarf
\end{minipage}

\begin{minipage}[b]{0.19\linewidth}
\centering
\centerline{\includegraphics[width=1.01\linewidth]{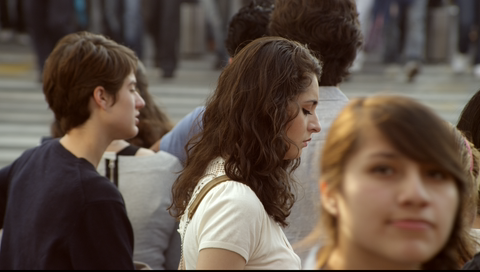}}
(p) Cross Walk
\end{minipage}
\begin{minipage}[b]{0.19\linewidth}
\centering
\centerline{\includegraphics[width=1.01\linewidth]{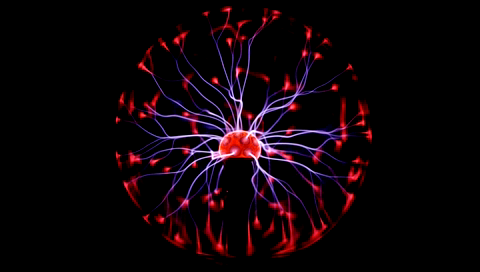}}
(q) Plasma
\end{minipage}
\begin{minipage}[b]{0.19\linewidth}
\centering
\centerline{\includegraphics[width=1.01\linewidth]{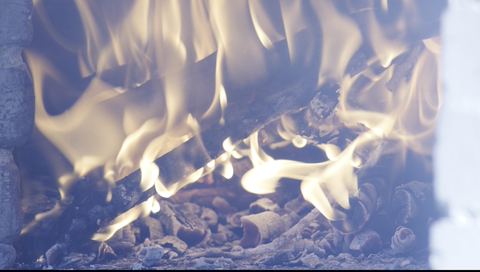}}
(r) Firewood
\end{minipage}
\begin{minipage}[b]{0.19\linewidth}
\centering
\centerline{\includegraphics[width=1.01\linewidth]{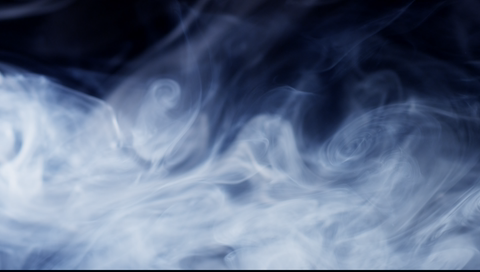}}
(s) Smoke
\end{minipage}
\begin{minipage}[b]{0.19\linewidth}
\centering
\centerline{\includegraphics[width=1.01\linewidth]{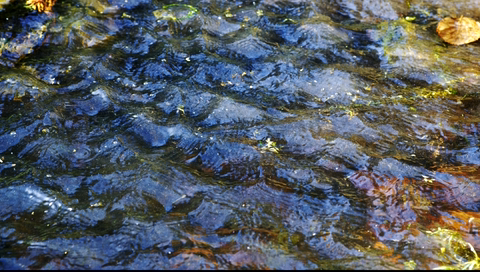}}
(t) Water
\end{minipage}
\caption{Sample frames of 20 example sequences from the BVI-DVC database.}
\label{fig:frame}
\end{figure*}

Two hundred source sequences were carefully selected from public video databases, including 69 sequences from the Videvo Free Stock Video Footage set \cite{videvo}, 37 from the IRIS32 Free 4K Footage set \cite{iris32}, 25 from the Harmonics database \cite{harmonics}, 19 from BVI-Texture \cite{papadopoulos2015video}, 10 from the  MCML 4K video quality database \cite{cheon2017subjective}, 7 from BVI-HFR \cite{mackin2018study},  7 from the SJTU 4K video database \cite{song2013sjtu},  6 from LIVE-Netflix \cite{bampis2017study,bampis2018towards}, 6 from the Mitch Martinez Free 4K Stock Footage set \cite{mitch}, 5 from the Dareful Free 4K Stock Video data set \cite{dareful}, 3 from MCL-V \cite{lin2015mcl}, 2 from  MCL-JCV \cite{wang2016mcl}, 2 from Netflix Chimera \cite{netflix}, 1 from the TUM HD databases \cite{keimel2012tum}, and 1 from the Ultra Video Group-Tampere University database \cite{tampere}. These sequences contain natural scenes and objects \cite{nah2019ntireData}, e.g. mountains, oceans, animals, grass, trees, countryside, city streets, towns, buildings, institutes, facilities, parks, marketplaces, historical places, vehicles and colorful textured fabrics. Different texture types such as static texture, dynamic texture\footref{fn:texture}, structure content and luminance-plain content are also included.

All these sequences are progressive-scanned at a spatial resolution of 3840$\times$2160, with frame rates ranging from 24 fps to 120 fps, a bit depth of 10 bit, and in YCbCr 4:2:0  format. All are truncated to 64 frames without scene cuts, using the segmentation method described in \cite{moss2015optimal}. To further increase data diversity and provide data augmentation, the 200 video clips were spatially down-sample to 1920$\times$1080, 960$\times$540 and 480$\times$270 using a Lanczos filter of order 3. This results in 800 sequences at four different resolutions. Fig. \ref{fig:frame} shows the sample frames of twenty example sequences. The primary features of this database are summarised in Table \ref{tab:feature} alongside those for the other seven databases \cite{russakovsky2015imagenet,Agustsson_2017_CVPR_Workshops,martin2001database,xue2019video,liu2017robust,nah2019ntireData,soomro2012ucf101} mentioned above.

\begin{figure}[h]
\centering
\scriptsize
\centering
\begin{minipage}[b]{0.485\linewidth}
\centering
\centerline{\includegraphics[width=1.08\linewidth]{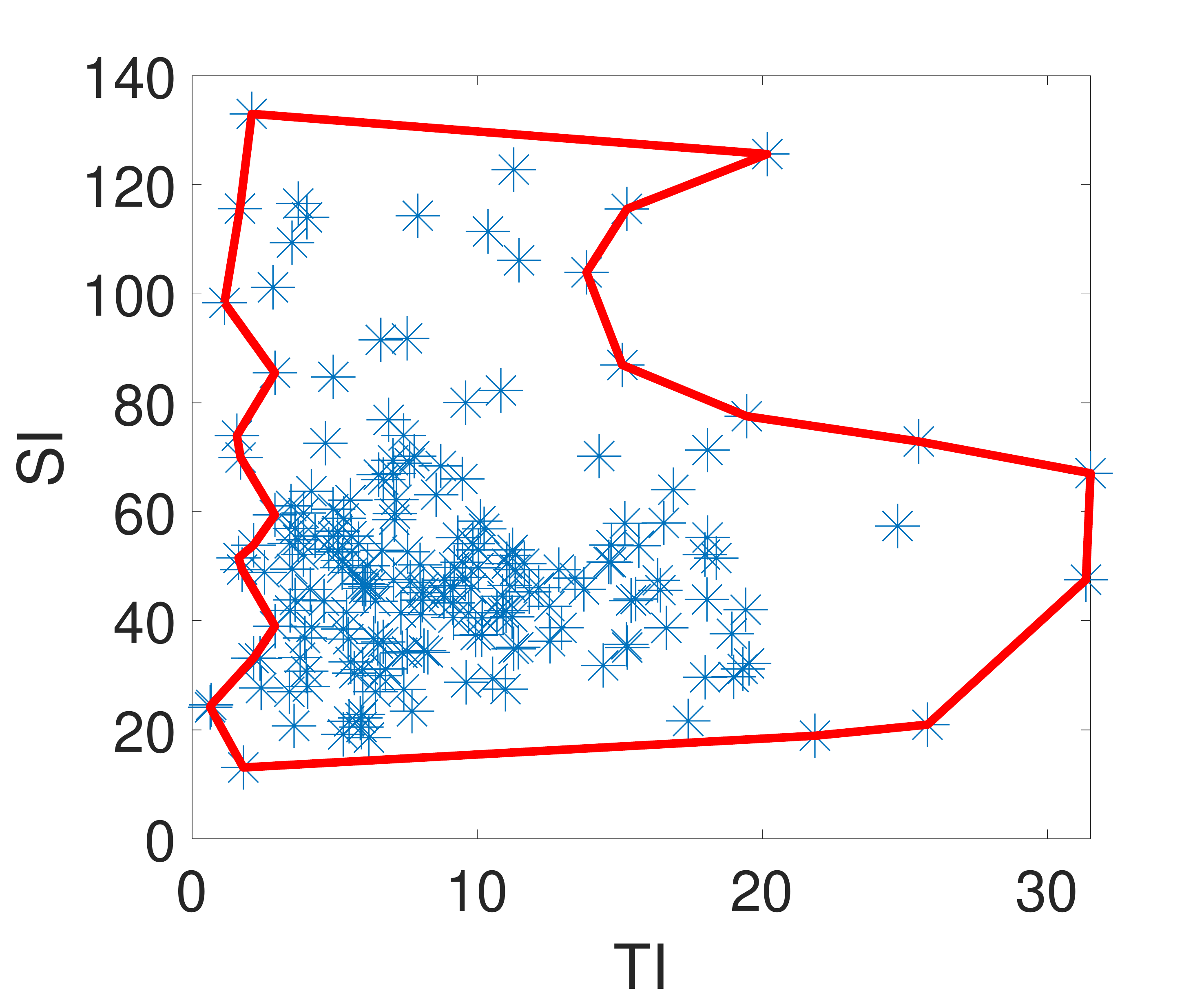}}
(a) SI vs. TI
\end{minipage}
\begin{minipage}[b]{0.485\linewidth}
\centering
\centerline{\includegraphics[width=1.08\linewidth]{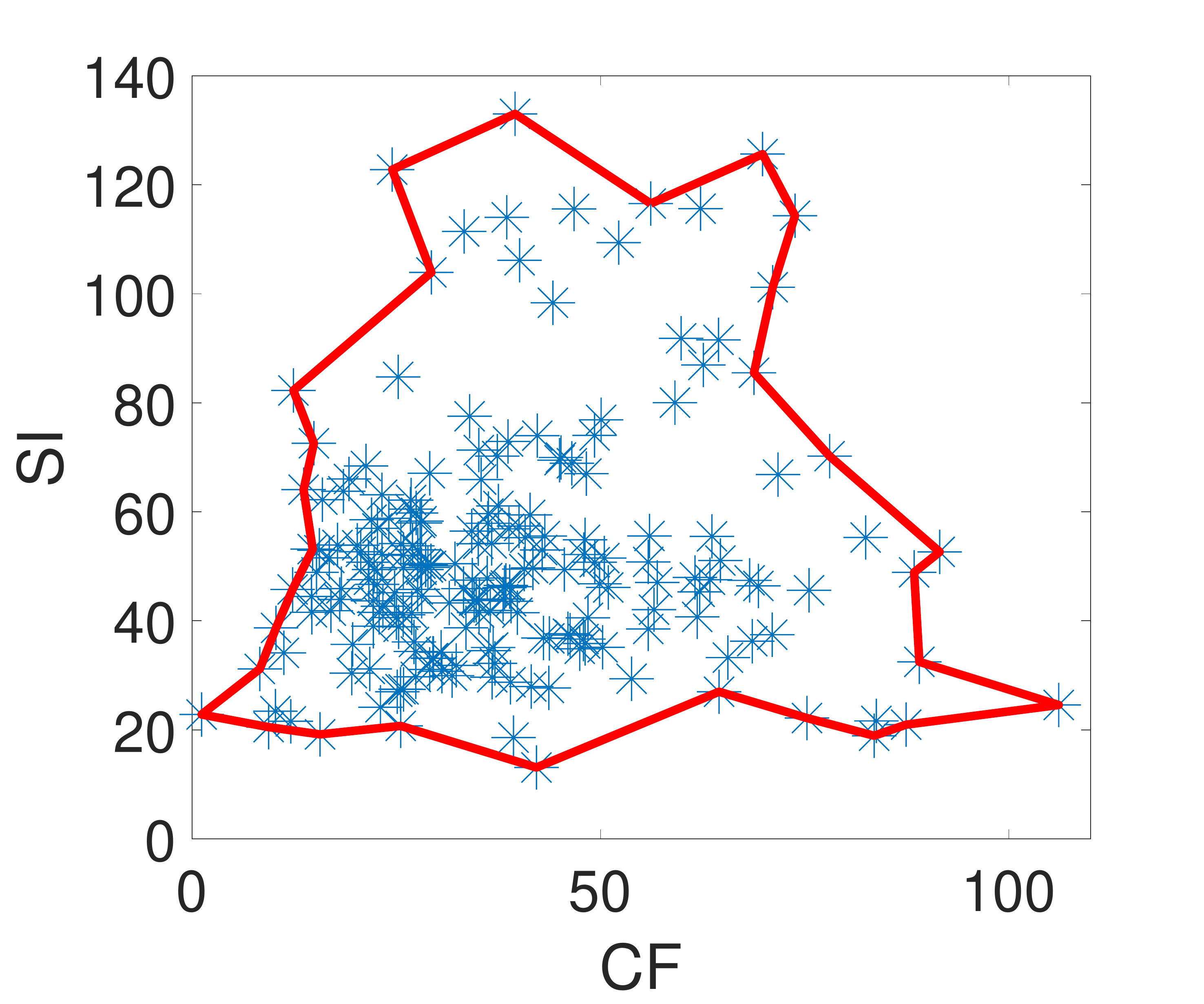}}
(b) SI vs. CF
\end{minipage}
\caption{Scatter plots of three video features for 200 UHD source sequences in the BVI-DVC database.}
\label{fig:feture}
\end{figure}

Three low-level video features of all 200 UHD source sequences, spatial information (SI), temporal information (TI) and colourfulness (CF) have been also calculated and plotted in Fig. \ref{fig:feture}. The definitions of these feature can be found in \cite{winkler2012analysis,mackin2018study}. It can be noted that the BVI-DVC database has a relatively wide coverage for these three video features, which indicates the diversity of the proposed database.

\section{Experiments}

In order to evaluate the training effectiveness of the proposed BVI-DVC database in the context of video compression, ten network architectures \cite{dong2015image,dong2016accelerating,kim2016accurate,tai2017image,lim2017enhanced,ledig2017photo,wang2018esrgan,zhang2018image,zhang2018residual,ma2019perceptually,JVET-J0031} were employed in conjunction with four CNN-based coding modules: post processing (PP), in-loop filtering (ILF),  spatial resolution adaptation (SRA) and effective bit depth adaptation (EBDA). These four coding modules were selected since they have been demonstrated to offer significant coding gains over standardised video codecs compared to other tools (e.g. inter prediction, entropy coding, etc.) and also outperform existing end-to-end solutions. Compared to end-to-end image coding  architectures, they are also more amenable to integration into standard codecs for practical applications.

In terms of benchmarking databases, based on the limited time and resource available, we have chosen one image database (DIV2K) and two video databases (REDS and CD) to compare with BVI-DVC. The DIV2K is selected because it has been used for training CNN models in multiple JVET contributions and many other CNN-based coding algorithms. Comparing to other video databases such as BSDS, Vimeo and UCF101, REDS and CD contain relatively higher resolution content and have been employed for training successful CNN-based super-resolution approaches.

\subsection{Coding Modules}

\subsubsection{Coding Module 1 (Post Processing - PP)}

\begin{figure}[htbp]
\centering
\includegraphics[width=1\linewidth]{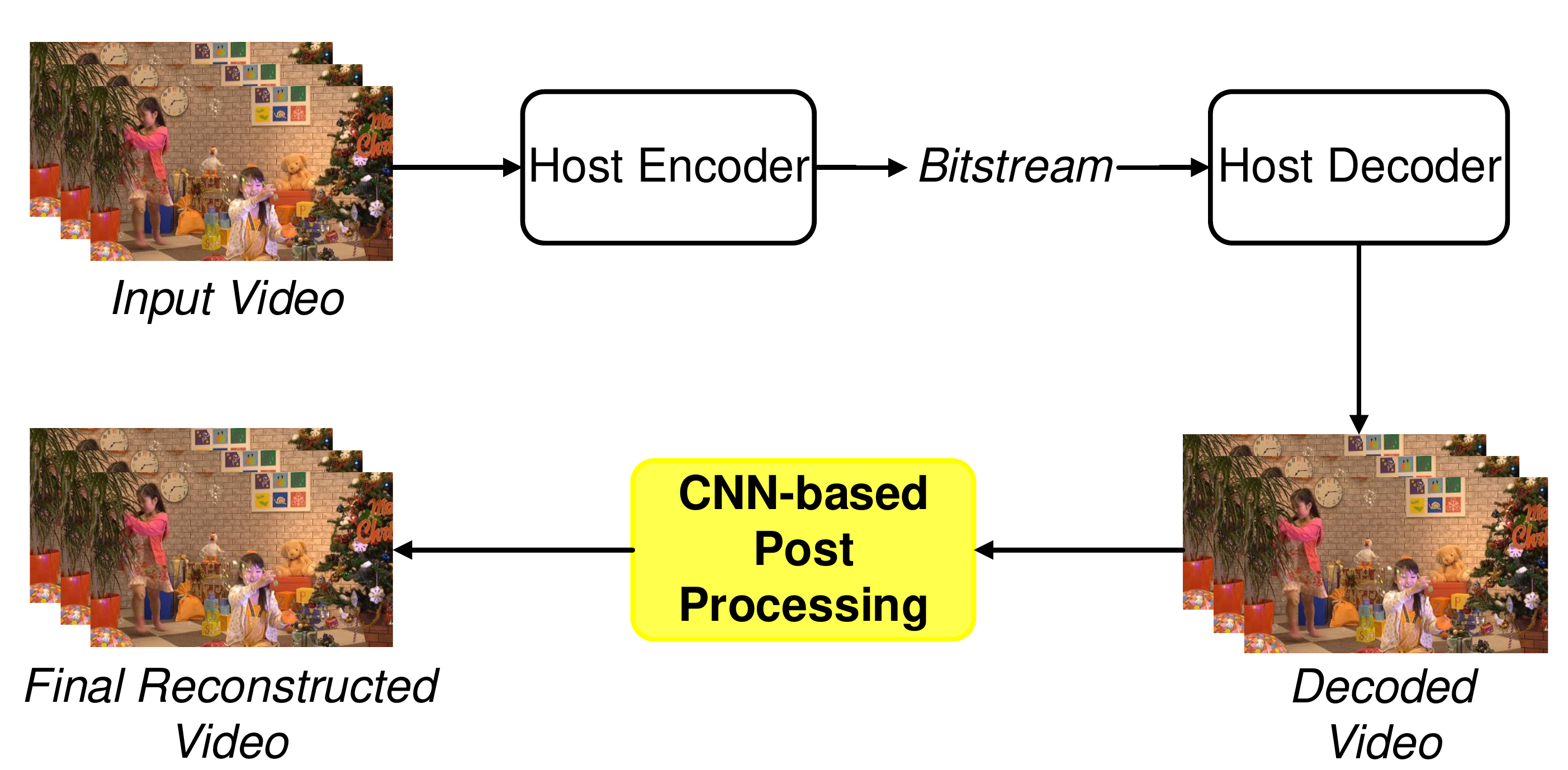}
\caption{Coding workflow with a CNN-based post processing module.}
\label{fig:pp}
\end{figure}

The coding workflow for post processing (PP) is illustrated in Fig. \ref{fig:pp}. PP is commonly applied at the decoder, on the reconstructed video frames, to reduce compression artefacts and enhance video quality. When a CNN-based approach is employed, the network takes each decoded frame as input and outputs the final reconstructed frame with the same format. Notable examples of employing CNN-based post processing for video coding can be found in \cite{li2017cnn,zhao2019cnn}.

\subsubsection{Coding Module 2 (In-loop Filtering - ILF)} 

\begin{figure}[htbp]
\centering
\includegraphics[width=1\linewidth]{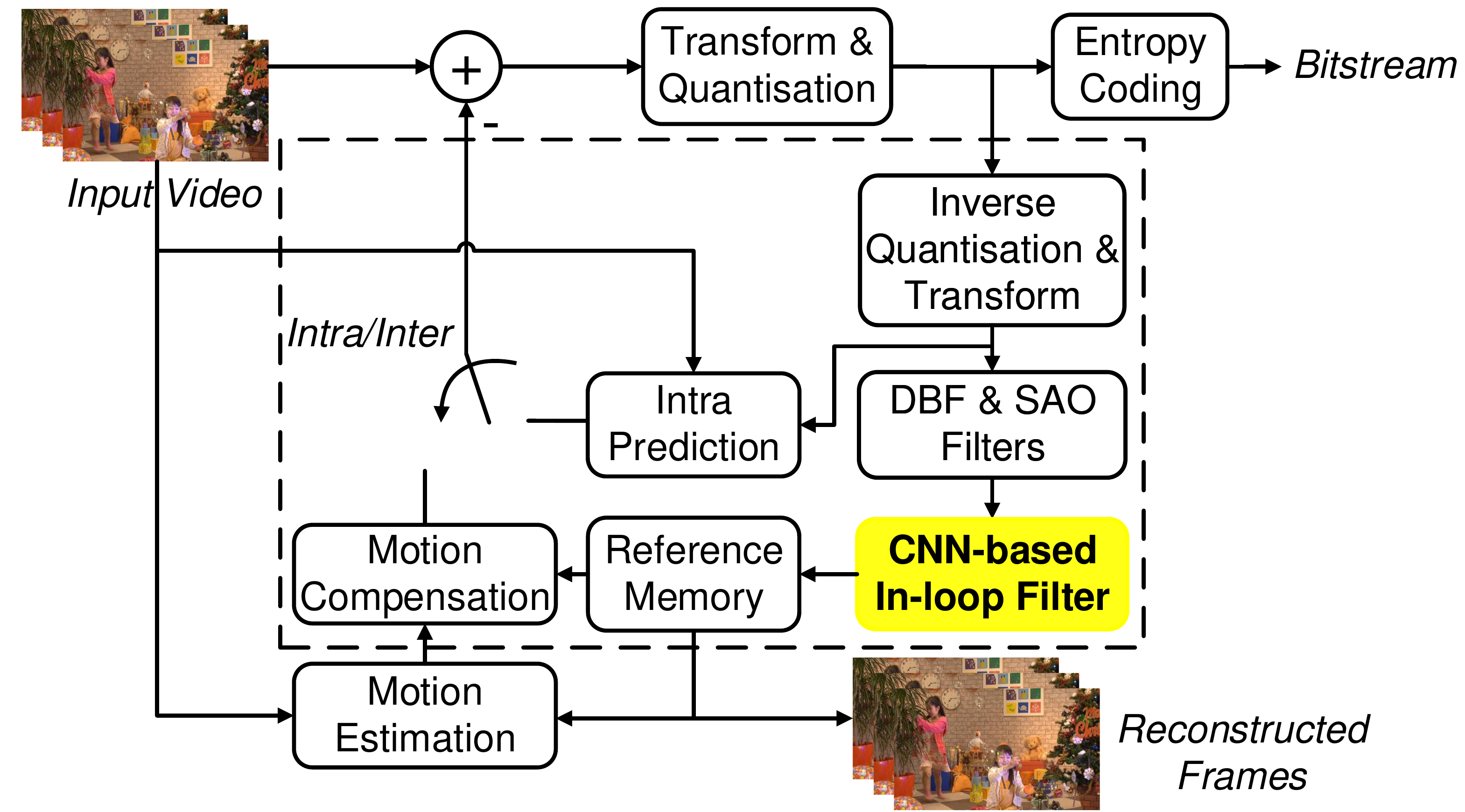}
\caption{Coding workflow with a CNN-based in-loop filtering module (the modules in the dashed box form the corresponding decoder). Here, DBF stands for de-blocking filtering, and SAO is denoted as sample adaptive offset.}
\label{fig:ilf}
\end{figure}

In-loop filtering applies processing at both the encoder and the decoder on the reconstructed frames, and the output can be used as reference for further encoding/decoding. An encoder architecture with a CNN-based ILF module is shown in Fig. \ref{fig:ilf}. The input and the output of the CNN-based ILF are the same as those for PP \cite{li2019deep,jia2019content}.

\subsubsection{Coding Module 3 (Spatial Resolution Adaptation - SRA)}  
\begin{figure}[htbp]
\centering
\includegraphics[width=1\linewidth]{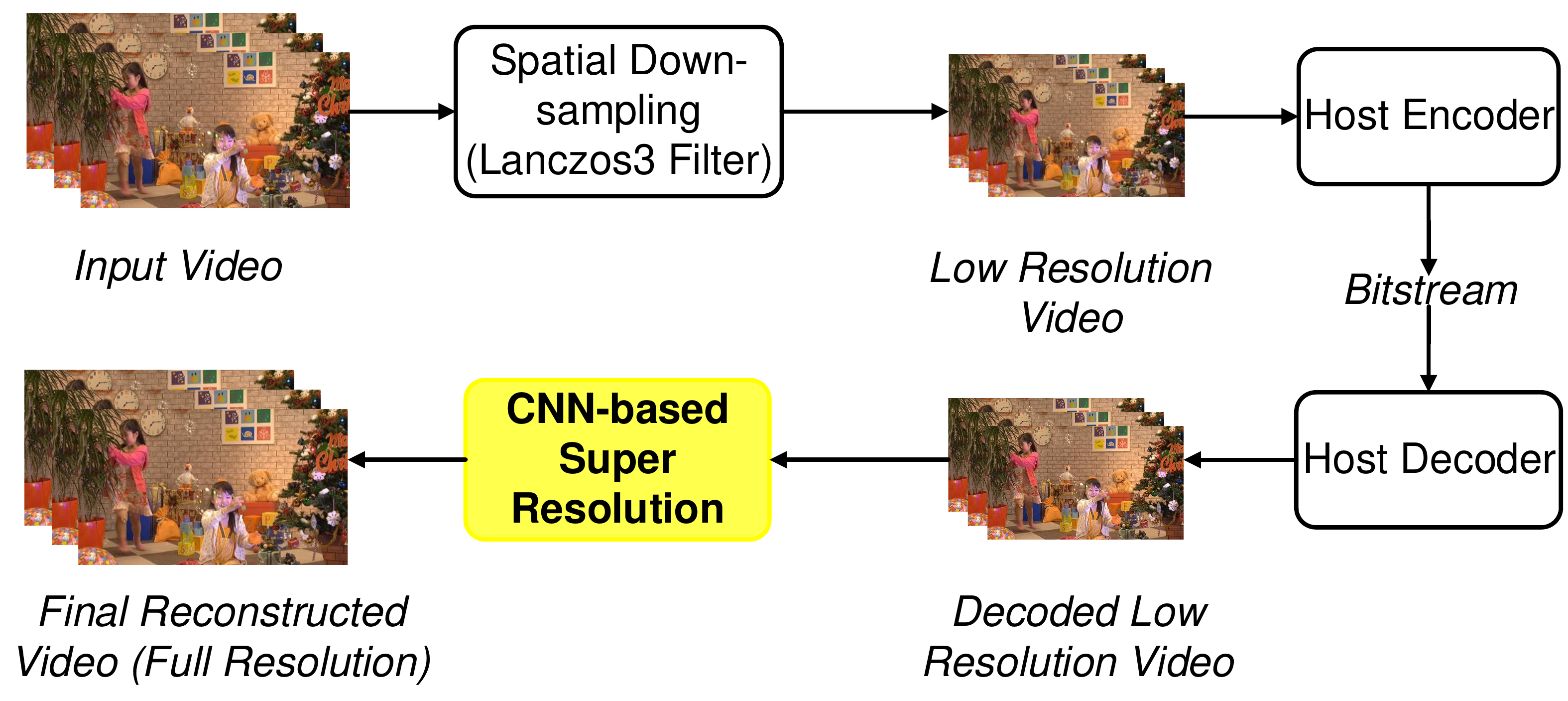}
\caption{Coding workflow with a CNN-based spatial resolution adaptation module.}
\label{fig:sra}
\end{figure}

CNN-based spatial resolution adaptation (SRA) down-samples the spatial resolution of the original video frames for encoding, and reconstructs the full resolution during decoding through CNN-based super-resolution. This approach can be applied at Coding Tree Unit level \cite{lin2018convolutional} or to the whole frame. Here we only implemented frame-level SRA \cite{afonso2019video}, as shown in Fig. \ref{fig:sra}. In this case, the original video frames are spatially down-sampled by a fixed factor of 2, using the Lanczos3 filter. The CNN-based super-resolution module processes the compressed and down-sampled video frames at the decoder to generate full resolution reconstructed frames. It is noted that a nearest neighbour filter is firstly applied to the reconstructed down-sampled video frame before CNN operation \cite{zhang2019vistra2}.

\subsubsection{Coding Module 4 (Effective Bit Depth Adaptation - EBDA)} 
\begin{figure}[htbp]
\centering
\includegraphics[width=1\linewidth]{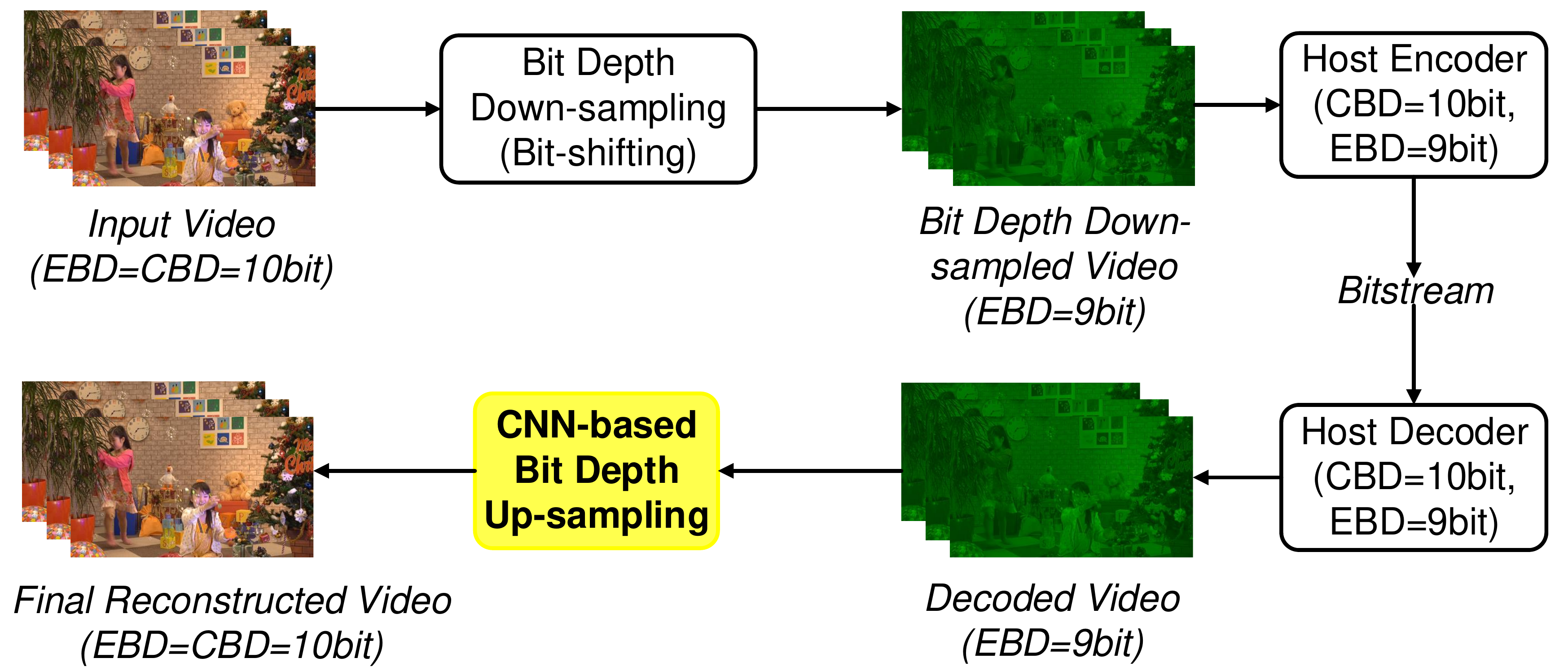}
\caption{Coding workflow with a CNN-based effective bit depth adaptation module.}
\label{fig:ebda}
\end{figure}

Similar to the case for spatial resolution, bit depth can also be adapted during encoding in order to achieve improved coding efficiency. Here Effective Bit Depth (EBD) is defined as the actual bit depth used to represent the video content, which may be different from the Coding Bit Depth (CBD) that represents the pixel bit depth, e.g. \textit{InternalBitDepth} in HEVC reference encoders. This process is demonstrated in Fig. \ref{fig:ebda}. In this paper, we have fixed CBD at 10 bits, the same as in the Main10 profile of HEVC, and only down-sampled the original frames by 1 bit through bit-shifting. At the decoder, the bit depth of decoded frames is up-sampled to 10 bits using a CNN-based approach. More information about EBDA can be found in \cite{zhang2019BD}.

\subsection{Employed CNN Models}
\label{subsec:CNN}
For these four coding modules, ten popular network architectures have been implemented for evaluation. These include five with residual blocks, two with residual dense blocks, and three without any residual block structure. Most of these network structures were initially designed for super-resolution processing or image enhancement, and some  have been employed in CNN-based coding approaches as described in Section \ref{sec:review}. Their primary features are briefly described below:

\begin{itemize}
    \item \textbf{SRCNN} \cite{dong2015image} (trained by mean-squared-error (MSE) loss) is the first CNN model designed for single image super resolution (SISR). It employs a simple network structure with only 3 convolutional layers.
    \item \textbf{FSRCNN} \cite{dong2016accelerating} (trained by MSE loss) was also developed for SISR, containing 8 convolutional layers with various kernel sizes. 
    \item \textbf{VDSR} \cite{kim2016accurate} (trained by MSE loss) contains 20 convolutional layers employing global residual learning to achieve enhanced performance. However it does not employ residual blocks  \cite{he2016deep}, which may lead to unstable training and evaluation performance \cite{tai2017image}.
    \item \textbf{SRResNet} \cite{ledig2017photo} (trained by MSE loss) was the first network structure with residual blocks designed for SISR, improving the overall performance and stability of the network.
    \item \textbf{DRRN} \cite{tai2017image} (trained by MSE loss) employs a recursive structure and also contains residual blocks. 
    \item \textbf{EDSR} \cite{lim2017enhanced} (trained by $\ell 1$ loss) significantly increases the number of feature maps (256) for the convolutional layers in each residual block. It offers improved overall performance but with much higher computational complexity.
    \item \textbf{RDN} \cite{zhang2018residual} (trained by $\ell 1$ loss) was the first network architecture to combine residual block and dense connections \cite{huang2017densely} for SISR. 
    \item \textbf{ESRResNet} \cite{wang2018esrgan} (trained by $\ell 1$ loss) enhances SRResNet by combining residual blocks with dense connections, and employs residual learning at multiple levels. It also removes the batch normalisation (BN) layer used in SRResNet to further stabilise training and reduce artefacts. 
    \item \textbf{RCAN} \cite{zhang2018image} (trained by $\ell 1$ loss) incorporates a channel attention (CA) scheme in the CNN, which better recovers high frequency texture details. 
    \item \textbf{MSRResNet} \cite{ma2019perceptually} (trained by $\ell 1$ loss) modified SRResNet by removing the BN layers for all the residual blocks. This network structure has been employed in several CNN-based coding algorithms and has been reported to offer significant coding gains \cite{zhang2019vistra2}.
\end{itemize}

In this experiment, we have employed identical architectures for these ten networks, as reported in their original publications, and only modify the input and output interfaces in order to process content in the appropriate format. The input of all CNNs employed is a 96$\times$96 YCbCr 4:4:4 colour image, while the output targets the corresponding original image block with the same size. The input block can be either compressed (for PP and ILF), compressed and EBD down-sampled (for EBDA) or compressed and spatial resolution re-sampled (for SRA - a nearest neighbour filter is applied before CNN processing). The same loss functions have been used as in the corresponding literature. All these ten networks have been re-implemented using the TensorFlow framework (version 1.8.0).

\subsection{Training Data}

Three existing image and video databases are selected to benchmark the training effectiveness of BVI-DVC, including DIV2K \cite{Agustsson_2017_CVPR_Workshops}, REDS \cite{nah2019ntireData} and CD \cite{liu2017robust}. All the original images or videos in each database were first spatially down-sampled by a factor of 2 using a Lanczos3 filter or down-sampled by 1 bit through bit-shifting. The original content (for training PP and ILF CNNs), together with spatially down-sampled clips (for training SRA CNNs) and bit depth reduced sequences (for training EBDA CNNs) were then compressed by the HEVC Test Model (HM 16.20) based on the JVET Common Test Conditions (CTC) \cite{jvetctc} using the Random Access configuration (Main10 profile) with four base QP (quantisation parameter) values: 22, 27, 32 and 37\footnote{Here results with four QP values are generated due to the limited time and resource given. During evaluation, if the base QP is different from these four, the CNN model for the closest QP value will be used.} (a fixed QP offset of -6 is applied for both spatially and bit depth down-sampled cases as in \cite{afonso2017low}). This results in three training input content groups for every database, each of which contains four QP sub-groups. For the input content group with reduced spatial resolution, a nearest neighbour filter was applied to obtain video frames with the same size as the original content.

For each input group and QP sub-group, the video frames of all reconstructed sequences and their original counterparts were randomly selected (with the same spatial and temporal sampling rates) and split into 96$\times$96 image blocks, which were then converted to YCbCr 4:4:4 format. Block rotation was also applied here for data augmentation.

\subsection{Network Training and Evaluation}

\begin{table*}[ht]
\centering
\caption{Evaluation results for PP coding module for ten tested network architectures and four different training databases. Values indicate the average BD-rate (\%) for all nineteen JVET CTC tested sequences assessed by PSNR or VMAF.}
\begin{tabular}{l | r | r |r |r|r|r|r|r}
\toprule
\multirow{2}{*}{CNN Model (PP)} & \multicolumn{2}{c|}{DIV2K \cite{Agustsson_2017_CVPR_Workshops}} &
\multicolumn{2}{c|}{REDS \cite{nah2019ntireData}}&\multicolumn{2}{c|}{CD \cite{liu2017robust}}&\multicolumn{2}{c}{BVI-DVC}\\
\cmidrule{2-9}
\centering
&   BD-rate &    BD-rate &BD-rate& BD-rate &BD-rate &BD-rate &BD-rate&BD-rate\\
&(PSNR)&(VMAF) &(PSNR)& (VMAF) &(PSNR)&  (VMAF)&(PSNR)& (VMAF)\\
\midrule \midrule
SRCNN \cite{dong2015image}&0.4 & -2.8 &2.5 & -3.8 &11.0&-6.2&\textbf{-1.9}&\textbf{-7.4}\\
\midrule
FSRCNN \cite{dong2016accelerating}&0.7 & -2.6 &3.2 & -1.2 &24.4&2.6&\textbf{-1.6}&\textbf{-7.3}\\
\midrule
VDSR \cite{kim2016accurate}&0.3 & -2.9 &2.3 & -4.0 &3.1&-2.2&\textbf{-1.9}&\textbf{-7.6}\\
\midrule
DRRN \cite{tai2017image}&-5.0 & -6.1 &-4.7 & -8.2 &0.4&-1.1&\textbf{-10.8}&\textbf{-14.9}\\
\midrule
EDSR \cite{lim2017enhanced}&-5.4 & -4.9 &-3.1 & -6.1 &-0.8&-6.5&\textbf{-10.0}&\textbf{-14.6}\\
\midrule
SRResNet \cite{ledig2017photo}&-5.3 & -5.4 &-4.0 & -9.0 &4.0&-3.8&\textbf{-9.8}&\textbf{-12.7}\\
\midrule
ESRResNet \cite{wang2018esrgan}&-6.9 & -6.7 &-6.1 & -9.4&-3.5&-9.1 &\textbf{-11.8}&\textbf{-17.7}\\
\midrule
RCAN \cite{zhang2018image}&-6.6 & -7.3 &-6.3 & -9.7 &-4.5&-11.0&\textbf{-12.1}&\textbf{-18.5}\\
\midrule
RDN \cite{zhang2018residual}&-7.0 & -7.2 &-6.9 & -10.6&-4.6&-10.9&\textbf{-12.2}&\textbf{-17.0} \\
\midrule
MSRResNet \cite{ma2019perceptually}&-6.4 & -6.5 &-5.3 & -9.2 &-2.6&-8.7&\textbf{-10.4}&\textbf{-14.2}\\
\bottomrule
\end{tabular}
\label{tab:PP}
\end{table*}

The training process was conducted using the following parameters: Adam optimisation \cite{kingma2014adam} with the following hyper-parameters:  $\beta_1$=0.9 and $\beta_2$=0.999; batch size of 16; 200 training epochs; learning rate (0.0001); weight decay of 0.1 for every 100 epochs. This generates 480 CNN models for 4 training databases, 3 input content groups (PP and ILF use the same CNN models), 4 QP sub-groups and 10 tested network architectures.

During the evaluation stage, for a specific coding module, the decoded video frames (already up-sampled to the same spatial resolution if needed) are firstly segmented into 96$\times$96 overlapping blocks with an overlap size of 4 pixels as CNN input (YCbCr 4:4:4 conversion). The output blocks are then aggregated following the same pattern and then converted to YCbCr 4:2:0 format to form the final reconstructed frame. 

\section{Results and Discussions}

Four different coding modules have been integrated into the HEVC (HM 16.20) reference software, and have been fully tested under JVET CTC  \cite{jvetctc} using the Random Access configuration (Main10 profile). Nineteen JVET-CTC SDR (standard dynamic range) video sequences from resolution classes A1, A2, B, C and D were employed as test content, none of which were included in any of the three training databases. It is noted that only class A1 and A2 (2160p) were used to evaluate SRA coding module, as it has been previously reported \cite{zhang2019vistra2} that for lower resolutions SRA may provide limited and inconsistent coding gains.

\begin{table*}[ht]
\centering
\caption{Evaluation results for ILF coding module for ten tested network architectures and four different databases. Each value indicates the average BD-rate (\%) for all nineteen JVET CTC tested sequences assessed by PSNR or VMAF.}
\begin{tabular}{l | r | r |r |r|r|r|r|r}
\toprule
\multirow{2}{*}{CNN Model (ILF)} & \multicolumn{2}{c|}{DIV2K \cite{Agustsson_2017_CVPR_Workshops}} &
\multicolumn{2}{c|}{REDS \cite{nah2019ntireData}}&\multicolumn{2}{c|}{CD \cite{liu2017robust}}&\multicolumn{2}{c}{BVI-DVC}\\
\cmidrule{2-9}
\centering
&   BD-rate &    BD-rate &BD-rate& BD-rate &BD-rate &BD-rate &BD-rate&BD-rate\\
&(PSNR)&(VMAF) &(PSNR)& (VMAF) &(PSNR)&  (VMAF)&(PSNR)& (VMAF)\\
\midrule \midrule
SRCNN \cite{dong2015image}&-0.2 & -2.6 &-0.6 & -2.5 &-0.1&-1.4&\textbf{-1.4}&\textbf{-8.5}\\
\midrule
FSRCNN \cite{dong2016accelerating}&-0.1 & -2.2 &-0.2 & -1.1 &0.0&-0.5&\textbf{-1.3}&\textbf{-8.1}\\
\midrule
VDSR \cite{kim2016accurate}&-1.0 & -1.1 &-0.7 & -2.7 &0.0&-0.5&\textbf{-2.2}&\textbf{-6.5}\\
\midrule
DRRN \cite{tai2017image}&-4.0 & -5.6 &-3.1 & -4.6 &0.4&-1.1&\textbf{-6.8}&\textbf{-11.0}\\
\midrule
EDSR \cite{lim2017enhanced}&-4.5 & -6.1 &-2.7 & -3.1 &-1.3&-4.0&\textbf{-5.9}&\textbf{-9.9}\\
\midrule
SRResNet \cite{ledig2017photo}&-5.1 & -8.6 &-2.9 & -4.0 &-1.1&-2.8&\textbf{-6.4}&\textbf{-10.6}\\
\midrule
ESRResNet \cite{wang2018esrgan}&-5.8 & -8.6 &-3.3 & -5.1&-2.5&-6.8 &\textbf{-7.3}&\textbf{-12.0}\\
\midrule
RCAN \cite{zhang2018image}&-5.4 & -8.5 &-6.3 & -9.7 &-3.0&-8.5&\textbf{-7.4}&\textbf{-11.4}\\
\midrule
RDN \cite{zhang2018residual}&-5.8 & -8.8 &-3.7 & -5.6&-3.0&-8.9&\textbf{-7.5}&\textbf{-11.8} \\
\midrule
MSRResNet \cite{ma2019perceptually}&-5.6 & -9.4 &-4.6 & -6.7 &-2.0&-6.5&\textbf{-6.4}&\textbf{-11.3}\\
\bottomrule
\end{tabular}
\label{tab:ILF}
\end{table*}

\begin{figure*}[ht]
\centering
\centering
\begin{minipage}[b]{0.45\linewidth}
\centering
\centerline{\includegraphics[width=1.0\linewidth]{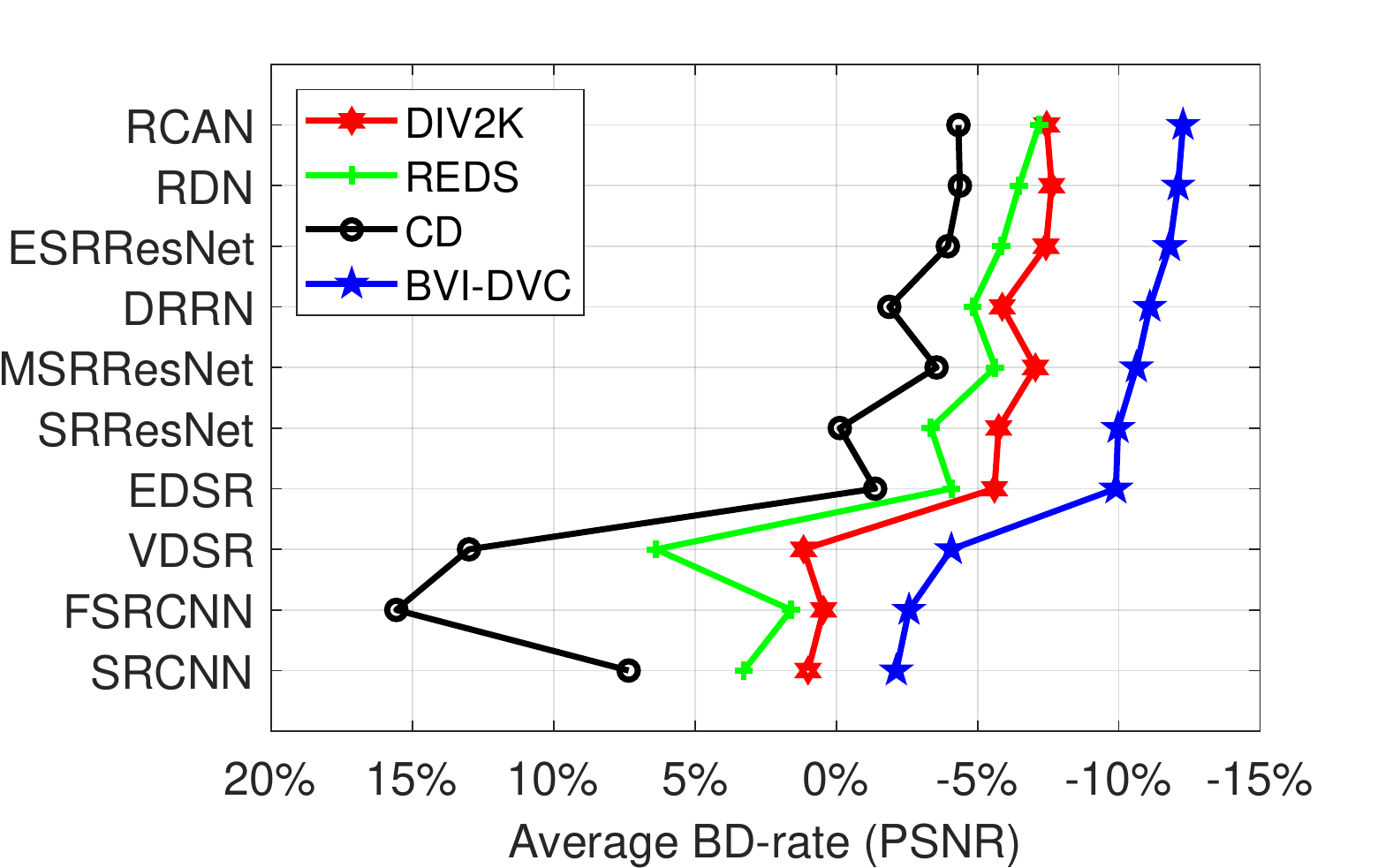}}
\end{minipage}
\begin{minipage}[b]{0.45\linewidth}
\centering
\centerline{\includegraphics[width=1.0\linewidth]{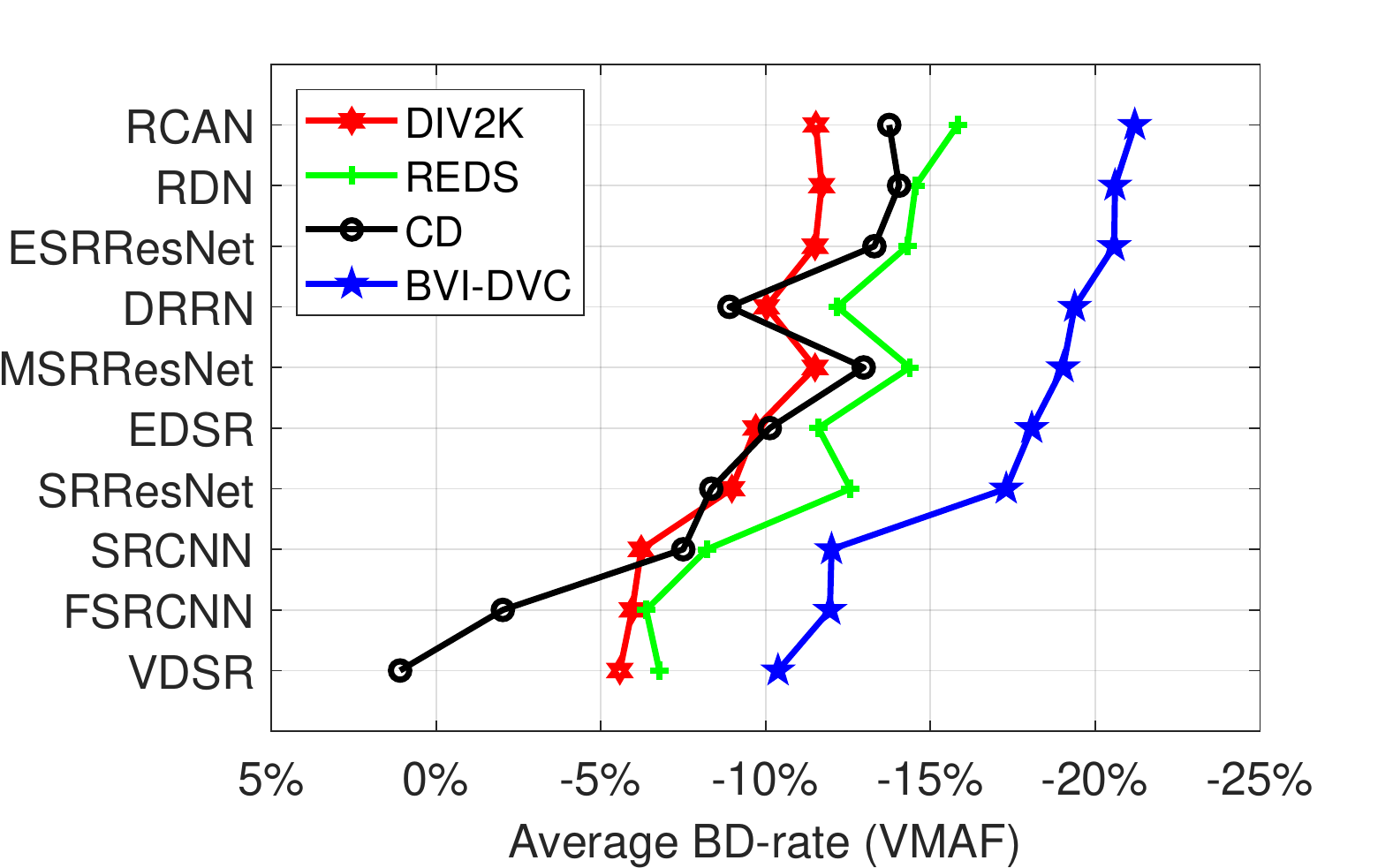}}
\end{minipage}
\caption{Average coding gains for four coding modules obtained using 10 commonly employed network architectures trained on four different databases: BVI-DVC, DIV2K \cite{Agustsson_2017_CVPR_Workshops}, REDS \cite{nah2019ntireData} and CD \cite{liu2017robust} All methods are integrated into HEVC Test Model (HM 16.20).}
\label{fig:bd}
\end{figure*}

\begin{figure*}[h]
\centering
\centering

\begin{minipage}[b]{0.45\linewidth}
\centering
\centerline{\includegraphics[width=1\linewidth]{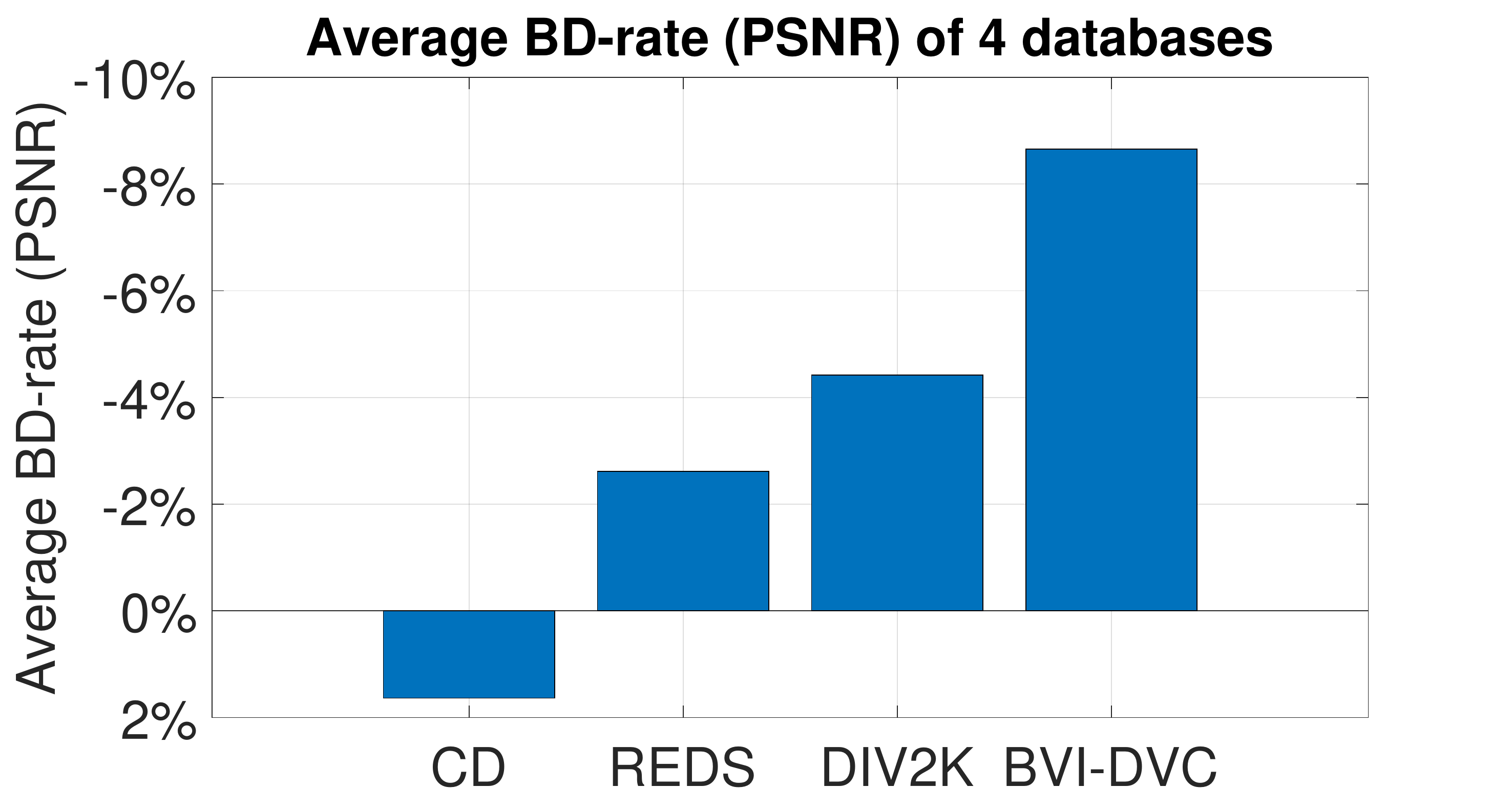}}
\end{minipage}
\begin{minipage}[b]{0.45\linewidth}
\centering
\centerline{\includegraphics[width=1\linewidth]{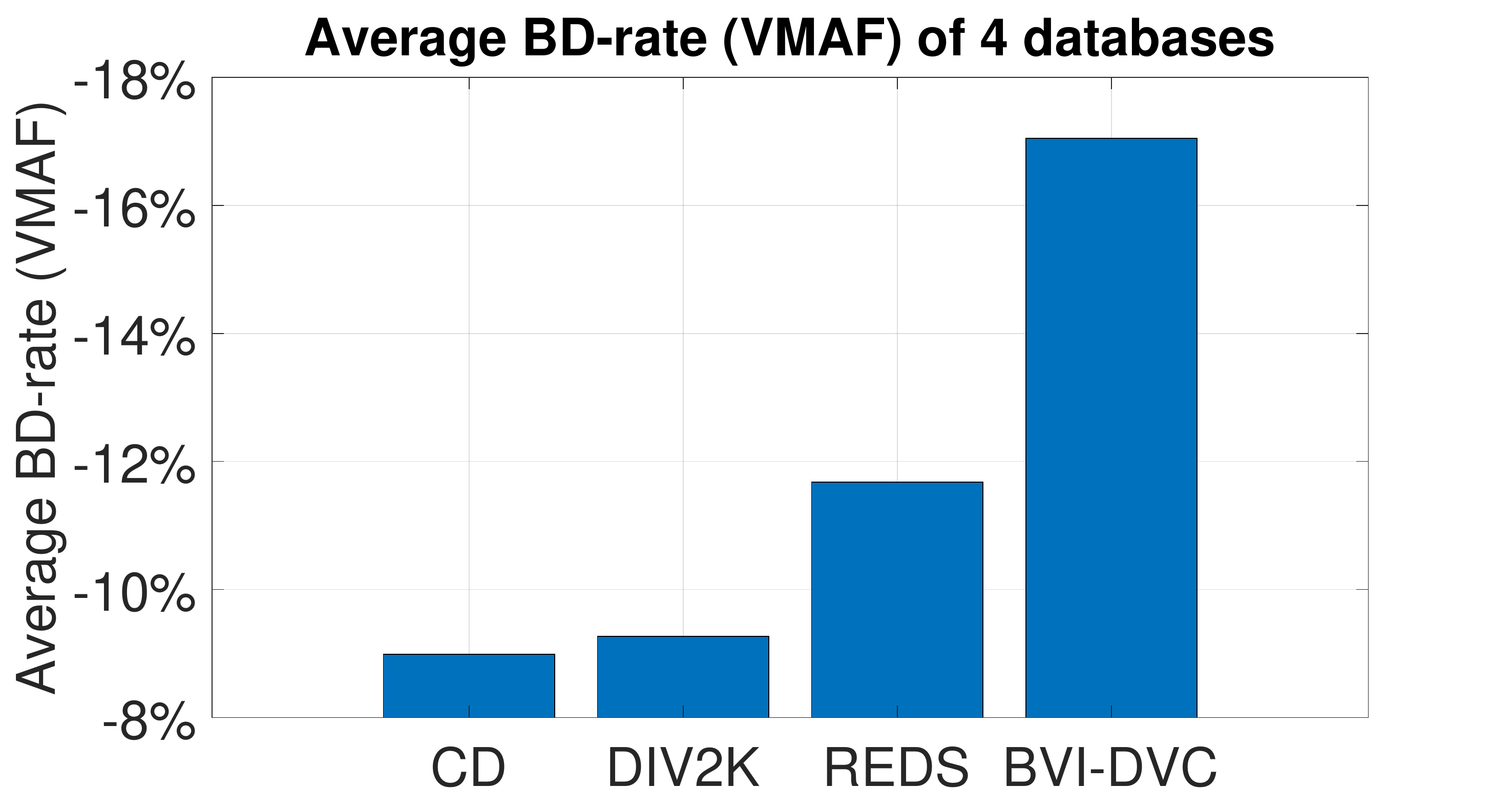}}
\end{minipage}
\caption{Average BD-rate (based on PSNR and VMAF) of four tested training databases for all the evaluated coding modules and CNN architectures.}
\label{fig:db}
\end{figure*}

The rate quality performance (coding performance) is benchmarked against the original HEVC HM 16.20, using Bj{\o}ntegaard Delta  measurement (BD-rate) \cite{BD} based on two quality metrics, Peak Signal-to-Noise  Ratio  (PSNR, Y-luminance channel only) and  Video  Multimethod  Assessment  Fusion (VMAF, version 0.6.1) \cite{li2016toward}. PSNR is the most commonly used assessment method for video compression, while VMAF is a machine learning based quality metric commercially used in streaming. VMAF combines multiple quality metrics and video features  using a Support Vector Machine (SVM) regressor. It has been reported to offer better correlation with subjective opinions \cite{zhang2018bvi}. BD-rate statistics indicate the overall bitrate savings across the tested QP range achieved by the test algorithm for the same video quality compared to the anchor approach. The average BD-rate values are reported in Table \ref{tab:PP}-\ref{tab:EBDA} for each evaluated training database, network architecture and coding module.

\subsection{Comparison of Databases}

\begin{table*}[ht]
\centering
\caption{Evaluation results for SRA coding module for ten tested network architectures  and four different databases.. Each value indicates the average BD-rate (\%) for all six UHD JVET CTC tested sequences assessed by PSNR or VMAF.}
\begin{tabular}{l | r | r |r |r|r|r|r|r}
\toprule
\multirow{2}{*}{CNN Model (SRA)} & \multicolumn{2}{c|}{DIV2K \cite{Agustsson_2017_CVPR_Workshops}} &
\multicolumn{2}{c|}{REDS \cite{nah2019ntireData}}&\multicolumn{2}{c|}{CD \cite{liu2017robust}}&\multicolumn{2}{c}{BVI-DVC}\\
\cmidrule{2-9}
\centering
&   BD-rate &    BD-rate &BD-rate& BD-rate &BD-rate &BD-rate &BD-rate&BD-rate\\
&(PSNR)&(VMAF) &(PSNR)& (VMAF) &(PSNR)& (VMAF)&(PSNR)& (VMAF)\\
\midrule \midrule
SRCNN \cite{dong2015image}&3.9 & -11.9 &8.6 & -19.6 &6.6&-12.4&\textbf{-3.1}&\textbf{-21.1}\\
\midrule
FSRCNN \cite{dong2016accelerating}&1.1 & -12.3&-0.3 & -18.0&9.9&-9.8&\textbf{-4.5}&\textbf{-20.9}\\
\midrule
VDSR \cite{kim2016accurate}&4.4 & -11.5 &4.3 & -15.9 &25.9&7.2&\textbf{-6.6}&\textbf{-18.3}\\
\midrule
DRRN \cite{tai2017image}&-8.5 & -17.6 &-7.8 & -26.1 &-7.2&-22.1&\textbf{-15.0}&\textbf{-33.2}\\
\midrule
EDSR \cite{lim2017enhanced}&-6.4 & -16.3 &-6.9 & -26.1 &-3.2&-20.4&\textbf{-13.4}&\textbf{-30.1}\\
\midrule
SRResNet \cite{ledig2017photo}&-6.7 & -11.5 &-7.0 & -28.1 &-5.5&-19.8&\textbf{-13.2}&\textbf{-30.0}\\
\midrule
ESRResNet \cite{wang2018esrgan}&-9.9 & -19.4 &-9.9 & -31.7&-7.8&-23.5 &\textbf{-16.1}&\textbf{-33.6}\\
\midrule
RCAN \cite{zhang2018image}&-10.2 & -19.3 &-10.9 & -32.2 &-8.4&-23.2&\textbf{-17.1}&\textbf{-35.1}\\
\midrule
RDN \cite{zhang2018residual}&-10.0 & -19.1 &-9.7 & -31.4&-8.4&-22.7&\textbf{-16.6}&\textbf{-34.5} \\
\midrule
MSRResNet \cite{ma2019perceptually}&-9.2 & -18.9 &-8.5 & -29.9 &-7.1&-22.8&\textbf{-14.6}&\textbf{-32.7}\\
\bottomrule
\end{tabular}
\label{tab:SRA}
\end{table*}

\begin{figure*}[t]
\centering
\centering

\begin{minipage}[b]{0.45\linewidth}
\centering
\centerline{\includegraphics[width=1.01\linewidth]{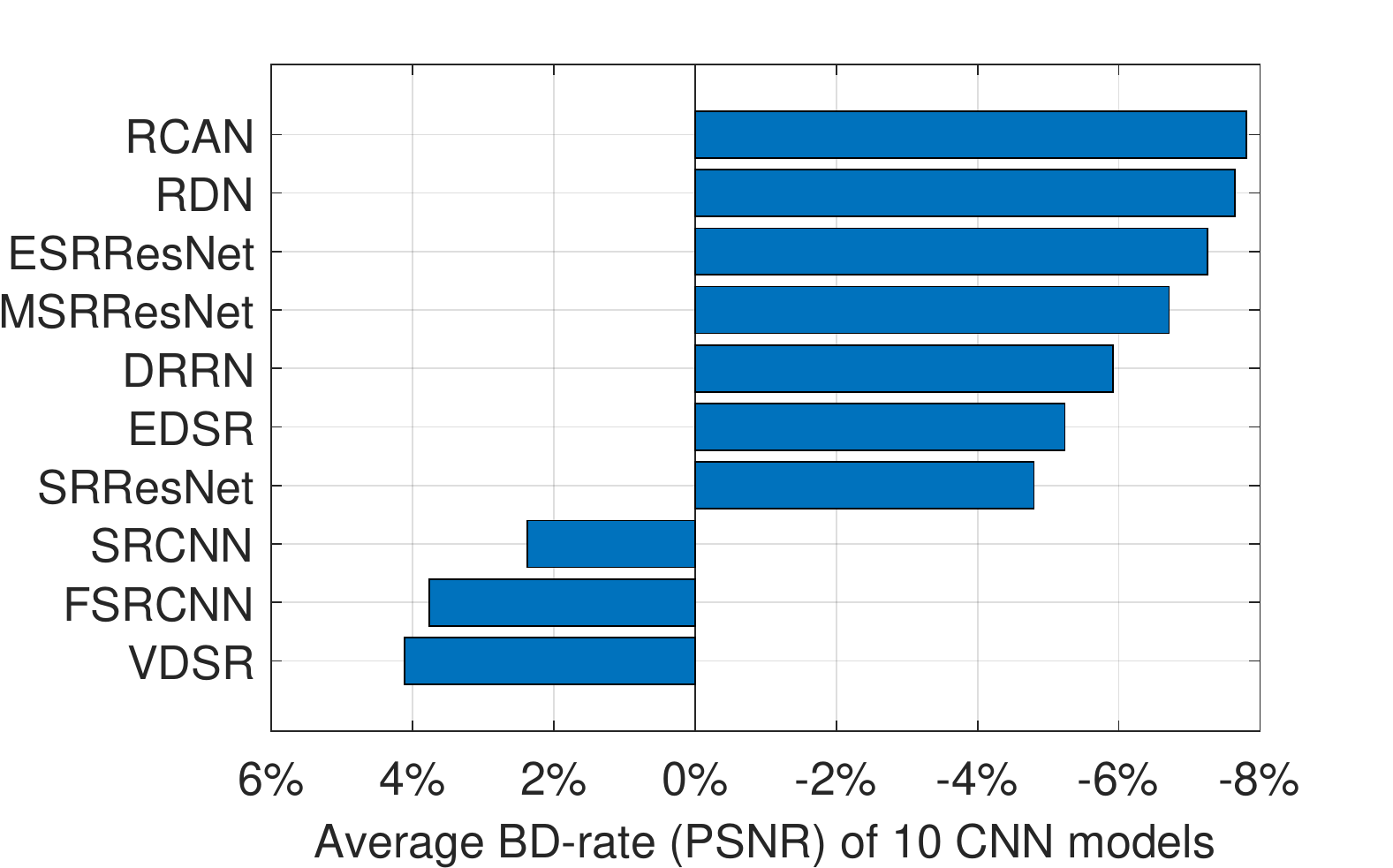}}
\end{minipage}
\begin{minipage}[b]{0.45\linewidth}
\centering
\centerline{\includegraphics[width=1.01\linewidth]{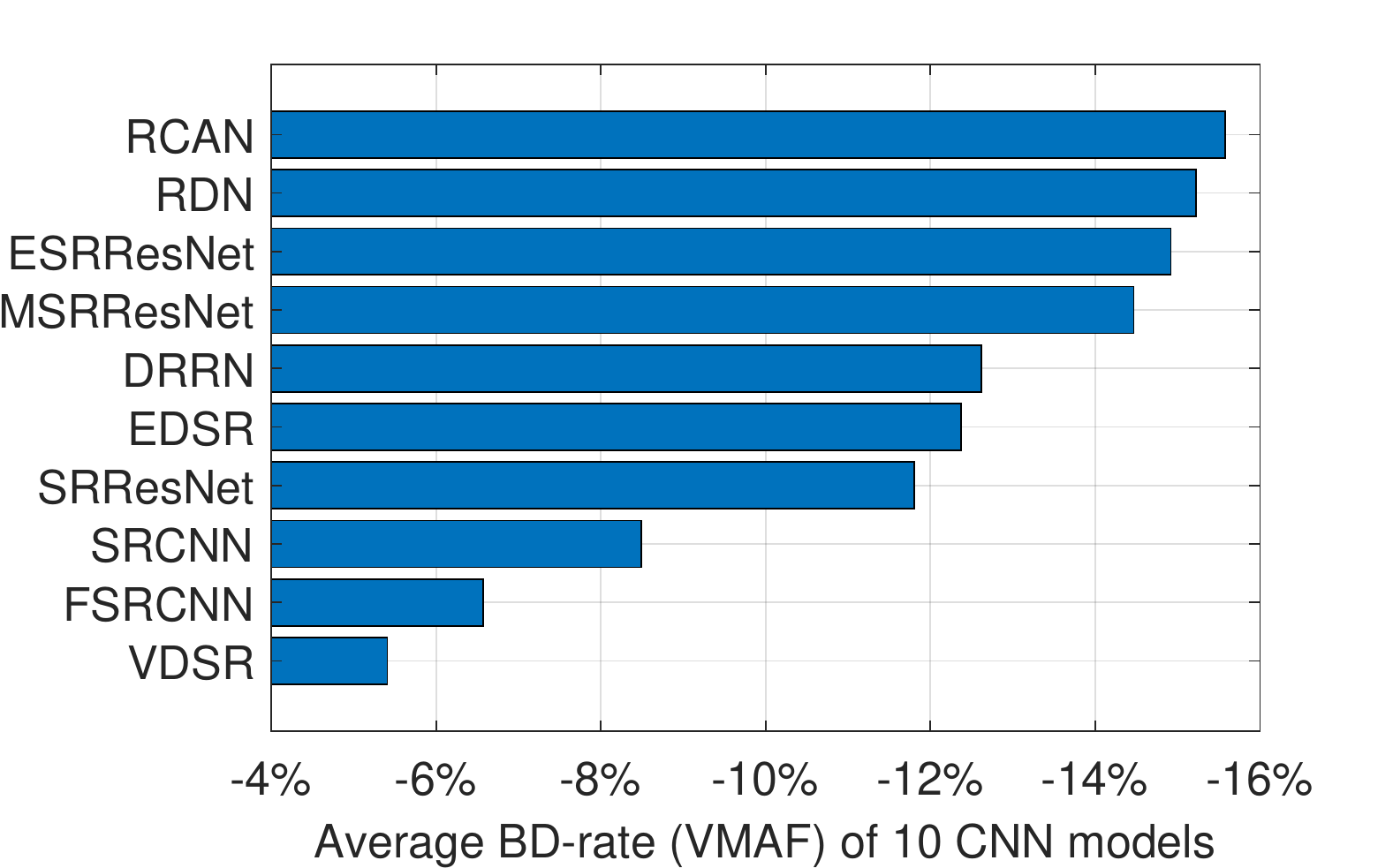}}
\end{minipage}
\caption{Average BD-rate (based on PSNR and VMAF) of 10 test network architectures for all coding modules and training databases.}
\label{fig:net}
\end{figure*}

It can be observed from Table \ref{tab:PP}-\ref{tab:EBDA} and Fig. {\ref{fig:bd}} that, for all tested network architectures and coding modules, the coding gains (in terms of average BD-rates for all tested sequences) achieved after training on the proposed BVI-DVC database are significantly greater than for the other three benchmark databases (DIV2K, REDS and CD) for both PSNR and VMAF quality metrics. This is reinforced by considering the mean (among ten networks and four coding modules) of all the average BD-rates for each database; Fig. \ref{fig:db} shows in excess of 4.2\% and 5.4\% additional bitrate savings obtained by using BVI-DVC compared to the other three databases based on the PSNR and VMAF quality metrics respectively. CD offers the worse overall performance, especially for results based on the assessment of PSNR.

\subsection{Comparison of Networks}

\begin{table*}[ht]
\centering
\caption{Evaluation results for EBDA coding module for ten tested network architectures and four different databases. Each value indicates the average BD-rate (\%) for all nineteen JVET CTC tested sequences assessed by PSNR or VMAF.}
\begin{tabular}{l | r | r |r |r|r|r|r|r}
\toprule
\multirow{2}{*}{CNN Model (EBDA)} & \multicolumn{2}{c|}{DIV2K \cite{Agustsson_2017_CVPR_Workshops}} &
\multicolumn{2}{c|}{REDS \cite{nah2019ntireData}}&\multicolumn{2}{c|}{CD \cite{liu2017robust}}&\multicolumn{2}{c}{BVI-DVC}\\
\cmidrule{2-9}
\centering
&   BD-rate &    BD-rate &BD-rate& BD-rate &BD-rate &BD-rate &BD-rate&BD-rate\\
&(PSNR)&(VMAF) &(PSNR)& (VMAF) &(PSNR)&  (VMAF)&(PSNR)& (VMAF)\\
\midrule \midrule
SRCNN \cite{dong2015image}&-0.1 & -7.6 &2.6 & -7.0 &11.9&-10.0&\textbf{-2.1}&\textbf{-11.0}\\
\midrule
FSRCNN \cite{dong2016accelerating}&0.1 & -6.7 &3.7 & -5.2&28.0&-0.4&\textbf{-2.9}&\textbf{-11.5}\\
\midrule
VDSR \cite{kim2016accurate}&0.93 & -6.8 &19.6 & -4.5 &23.0&-0.13&\textbf{-5.6}&\textbf{-9.1}\\
\midrule
DRRN \cite{tai2017image}&-6.0 & -10.8 &-3.7 & -9.8 &-1.1&-11.3&\textbf{-11.8}&\textbf{-18.4}\\
\midrule
EDSR \cite{lim2017enhanced}&-6.1 & -11.5 &-3.6 & -11.1 &-0.2&-9.6&\textbf{-10.3}&\textbf{-17.7}\\
\midrule
SRResNet \cite{ledig2017photo}&-5.9 & -10.4 &0.5 & -9.2 &2.1&-7.0&\textbf{-10.5}&\textbf{-15.9}\\
\midrule
ESRResNet \cite{wang2018esrgan}&-7.1 & -11.3 &-4.1 & -11.0&-2.0&-13.8 &\textbf{-12.0}&\textbf{-19.0}\\
\midrule
RCAN \cite{zhang2018image}&-7.6 & -11.0 &-5.2 & -11.7 &-1.4&-12.3&\textbf{-12.5}&\textbf{-19.8}\\
\midrule
RDN \cite{zhang2018residual}&-7.7 & -11.7 &-5.6 & -10.6&-1.5&-13.7&\textbf{-12.1}&\textbf{-19.1} \\
\midrule
MSRResNet \cite{ma2019perceptually}&-7.0 & -11.2 &-4.1 & -11.7 &-2.5&-13.9&\textbf{-11.1}&\textbf{-17.9}\\
\bottomrule
\end{tabular}
\label{tab:EBDA}
\end{table*}

We can also compare the ten evaluated network architectures under fair configurations (identical training and evaluation databases). The results in Table \ref{tab:PP}-\ref{tab:EBDA} are summarised, by taking the mean (among four training databases and four coding modules) of average BD-rate values for each network architecture, in Fig. \ref{fig:net}. It can be observed that RCAN, RDN, ESRResNet and MSRResNet offer better coding performance (for both PSNR and VMAF) than the other six evaluated network architectures. This is likely to be because of the residual block structure employed. The coding gains for VDSR, FSRCNN and SRCNN are relatively low comparing to other networks, exhibiting coding loss when PSNR is used to assess video quality especially when they are trained on the CD database (refer to Fig. \ref{fig:bd}). This may be due to their simple network architecture (FSRCNN and SRCNN) and the large number convolutional layers without residual learning structure (VDSR), which lead to less stable training and evaluation \cite{zhang2018residual,zhang2018image}.

\section{Conclusion}

This paper presents a new database (BVI-DVC) specifically designed for training CNN-based video compression  algorithms. With carefully selected sequences including diverse content, this database offers significantly improved training effectiveness compared to other commonly used image and video training databases. It has been evaluated for four different coding modules with ten typical CNN architectures. The BVI-DVC database reported is available online \footref{fn:db} for public testing. Its content diversity makes it a reliable training database, not just for CNN-based compression approaches, but also for other computer vision and image processing related tasks, such as image/video de-noising, video frame interpolation and super-resolution. Future work should focus on developing large training databases with more immersive video formats, including higher dynamic range, higher frame rates and higher spatial resolutions.



\section*{Acknowledgment}
The authors acknowledge funding from UK EPSRC (EP/L016656/1 and EP/M000885/1) and the NVIDIA GPU Seeding Grants.

\small
\bibliographystyle{IEEEtran}
\bibliography{refs}

\end{document}